%% file: 00.Main.tex
\RequirePackage{fix-cm}

\documentclass[smallextended]{svjour3}

\usepackage{cite}
\usepackage{graphicx}
\usepackage{listings}
\usepackage{hyperref}
\usepackage[table,xcdraw]{xcolor}
\usepackage{booktabs,caption}
\usepackage[flushleft]{threeparttable}
\usepackage{pifont}
\usepackage{fancybox}

\begin{document}

\title{Serverless architecture efficiency: an exploratory study}


\author{Samuel Lavoie and Anthony Garant and Fabio Petrillo}
\institute{Department of Computer and Software Engineering, Ecole Polytechnique de Montreal, Quebec, Canada\\
            \email{samuel.lavoie.m@gmail.com,atgarant@gmail.com@polymtl.ca}
\and
            \\Université du Québec de Chicoutimi\\
            Department of Mathematics and Computer science\\
            555 boulevard de l'Université\\
            Chicoutimi, Canada\\
            \email{fabio@petrillo.com}
}

\date{Received: date / Accepted: date}

\maketitle

\input{01.Abstract}



\input{10.Introduction}
\input{20.Background}
\input{70.RelatedWork}
\input{30.StudyDesign}
\input{40.Results}
\input{50.Discussion}

\input{60.ThreatsOfValidity}
\input{80.Conclusion}


\bibliographystyle{spphys}
\bibliography{references}
\end{document}

%% file: 01.Abstract.tex
\abstract{Cloud service provider propose services to insensitive customers to use their platform. Different services can achieve the same result at different cost. In this paper, we study the efficiency of a serverless architecture for running highly parallelizable tasks to compare theses services in order to find the most efficient in term of performance and cost. More precisely, we look at the compute time and at the cost per task for a given task. The tasks studied is the count of the occurrence of a given word in a corpus. We compare the serverless architecture to the Apache Spark map reduce technique commonly used for this type of task. Using AWS Lambda for the serverless architecture and Amazon EMR for the Apache Spark map reduce, with similar compute power, we show that the serverless technique achieve comparable performance in term of compute time and cost. We observed that the lambda function is a great approach for real time computing, while EMR is preferable for task that require long compute time.}

%% file: 10.Introduction.tex

\section{\uppercase{Introduction}}
\label{sec:introduction}

\noindent Serverless architecture is gaining popularity due to its interesting pricing model. The growth in popularity of this model certainly raises concerns as it is used as a replacement for existing service architectures that are at the moment mature, somehow scalable and maintainable \cite{icac:8082018}.

Serverless architecture is an execution model where the provider dynamically manage the resources requested by an application \cite{Roberts:2016}. Despite the name, serverless computing still require servers. In fact, serverless computing can be seen as leasing a server for a very small amount of time before releasing it for another application to use it. Most modern application use serverless computing for CPU heavy task and dedicate smaller server that they can scale horizontally for the routing. Typically, theses smaller server are stateless so that they can send the CPU heavy tasks to the lambda functions asynchonously and continue with the execution until the functions return its results. This create a non blocking pipieline that is very efficient, very scallable and easily maintainable \cite{Doerrfeld:2017}. A big promoter of this architecture is Netflix \cite{Wootoon:2017}, proving that the leap of faith is actually realisable when done correctly.

Since its inception, consumers have discovered a growing number of use cases for the serverless architecture. The most notable use cases are automated backups, scheduled Cron jobs, processing uploaded object (ie: on S3), analyzing log or simply processing and arbitrarily payload. All theses tasks are known to be CPU extensive, but each one of them already had one or many solutions in place. The serverless pay-per-use business model is interesting for all of theses task. Considering that hosting and compute power is one of the biggest recurring cost for startup aside from workforce \cite{Afridi:2014}, it is interesting for them to know the tradeoff of each solution in order to make the most profitable choices. Given all of the possible services that can achieve the same results, it can be expensive to prototype them all to assert which service is the most suitable for a given problem. The objective of this paper is to analyze the tradeoff of using a serverless architecture or a MapReduce architecture to compute highly parallelizable tasks. Amazon Spark MapReduce on Elastic MapReduce is know to be to preferred method for this type of task \cite{Shi:2015}. However, it has also been shown by \cite{Jonas:2017} that lambda functions can achieve the same level of parallelization, while having a pay-per-use business model. In this work, we will try to look at the impact and the gain of a serverless architecture. We will use the Amazon Web Services solution, but we assume that our conclusions can be applied to other cloud services provider. 

The contribution of this paper is two folds:

\begin{enumerate}
\item Benchmark the efficiency of Amazon EMR and Amazon Lambda to perform CPU intensive task
\item Provide a guideline to inform a potential user which use cases are better for Amazon EMR or Amazon Lambda.
\end{enumerate}

We organized the paper as follows.
In Section 2, we will present the related work. In Section 3, we will present the methodology used to compute the results. In Section 4, we will present and discuss the results. In Section 5, we will present the threat to validity. In Section 6, we will present the conclusion.

%% file: 20.Background.tex
\section{\uppercase{Background}}
\label{sec:background}

\subsection{Serverless architecture and Amazon lambda}

Notably, a serverless architecture is more cost-efficient than owning or renting a cluster due to the periods of non-utilization that are not billed to the consumer. A serverless architecture also make the code development easier, since the multi-threading and the scaling is taken care of by the provider.

In counterparts, a serverless architecture suffer from some latency when a function has to be initiated after not being used for a long period of time \cite{Zimmerman:2017}. Also, for very high CPU intensive task, a serverless architecture might not be suited due to the limited resources allocated for each function. Finally, the consumer does not have the control of their server. This can make the deployment, the monitoring harder. This also constraint the user to a small subset of languages \cite{Zimmerman:2017}.

Even thought the serverless architectures have been architectured a couple of decenie ago \cite{Khan:2001}. It is only in 2014 that the first implementation appeared by Amazon Web services, with their lambda functions \cite{Khan:2001}. Since then, most cloud service provider offer serverless computing. For exemple, Microsoft Azure offers Azure functions \cite{Azure:2017} and Google offers Cloud functions \cite{Google:2017}.

\subsection{Apache Spark Map reduce and Amazon EMR}

Amazon EMR provides a managed MapReduce that they claim to be ``easy [to use], fast, and cost-effective to process vast amounts of data across dynamically scalable Amazon EC2 instances. Amazon EMR business model is a pay-per-time model. you pay a per-second rate for every second you use, with a one-minute minimum. For example, a 10-node cluster running for 10 hours costs the same as a 100-node cluster running for 1 hour. The hourly rate depends on the instance type used (e.g. standard, high cpu, high memory, high storage, etc), and billing is calculated down to the second and shows time in decimal form." \cite{AmazonEMR:2017} The Lambda functions business model is a pay-per-use model. You pay only for the compute time you consume - there is no charge when your code is not running \cite{AmazonLambdaHistory:2017}. The two services are very similar and achieve the same thing, but in the former, you reserve compute nodes and in the later, the service assign a node for you for the period of time of the execution only.

Apache Spark is an open-source, distributed processing system commonly used for big data task that would usually not fit in RAM. Apache Sparks uses in-memory caching and optimized execution. Coupled with Hadoop MapReduce, this solution is ideal for task with a lot of data to process.

%% file: 70.RelatedWork.tex
\section{\uppercase{Related work}}
\label{sec:relatedwork}

\subsection{Serverless architecture}

\noindent Being relatively recent, serverless computing is just starting to see get some attention by the research community. A lot of the research has been trying to define what is serverless computing   and   Function-as-as-Service (FaaS) \cite{Jonas:2017}\cite{Spillner:2017}\cite{Varghese:2017}.

Some work has been done to benchmark the performance of serverless architectures. It has been shown by \cite{Jonas:2017} that it is possible to build a model that is general enough to implement a number of distributed computing model such as Bulk synchronous parallel. In fact, they show that highly parallelizable operation, such as matrix multiplication, can be parallelized using lambda with very low bottleneck. More precisely, they built their own framework to serialize any highly parallelizable task and send them to an arbitrary number of workers on lambda. They show that the aggregates TFLOPS scale linearly with the number of workers. They also show that the read/write throughput scale also linearly with the number of workers. Finally, they break down the time taken for each phase of the lambda function. They show that the initialization bottleneck of the lambda is 15\%. In certain application this bottleneck can be considerable.

Malawski \textit{et al.} \cite{Malawski:2017} compared the serverless architecture with the HyperFlow architecture and evaluate their benchmark on AWS, Google Cloud and IBM OpenWhisk. They execute their benchmark on Mersenne Twister and Linpack. They compare the CPU resources allocated with respect to the memory allocated. They observe that the CPU resources allocated on AWS scale linearly with the memory allocated, where the CPU resources allocated does not scale linearly with the memory allocated on Google cloud framework. They also observe that AWS achieve over 30 GFLOPS where Google cloud tops at 17 GFLOPS. This difference is due to the hardware used on each platform. 

\subsection{MapReduce architecture}

MapReduce being more mature, more research have been done on the subject. Apache Spark is considered a good implementation of the MapReduce algorithm.

Hameeza \textit{et al.} \cite{Hameeza:2016} compared a traditional spark cluster to a spark cluster on the cloud. To compare both approach, they use HiBench  benchmark suite. The tasks are Aggregate, Baysian, Join, PageRank, Join, Scan, Sleep, Sort, TeraSort and WordCount. They show that running spark on the cloud is more effective than the traditional spark cluster in term of speed and throughput for all the tasks they benchmarked. Also, it shows less CPU utilization. 

Reyes-Ortiz \textit{et al.}  \cite{ReyesOrtiz:2015} compared two distributed computing framework. The first one is MPI/OpenMP on BeoWulf that high-performance oriented and exploits multi-machine/multi- core infrastructures. The second one is Apache Spark on Hadoop which targets iterative algorithms through in-memory computing. They benchmark on two machine learning algorithm: K-neirest neighbour and Pegasos SVM. They show that MPI/OpenMP outperform Spark in term of raw processing and performance, but Spark provide better data management and the possibility of dealing with other aspects such as node failure and data replication.

%% file: 30.StudyDesign.tex
\section{\uppercase{Study Design}}
\label{sec:methodology}

\noindent The benchmark will be done on the the word count task. The goal of this task is to count the occurrence of each word in a corpus. This task is highly parallelizable and can be seperated in N workers where N is the number of words in the corpus. The experiment will be executed on different number of worker in order to evaluate the time for completion of the task. The corpus will be a dump of enwiki taken the 2017/07/01 [source]. When unzipped, the dump is 56 GB. The format of the dump is xml. For this reason, the tags will be filtered out in a preprocessing task executed before the benchmark. Another preprocessing task will be to separate the file in smaller files. This step is done for practical reason, because we cannot put all the 56 GB of data in memory. All the experiment will be done using the Amazon Web Service cloud computing stack. The compute time does not take into account the pre-processing. It takes into account the data processing time, the data streaming time and the startup time for each technique. For the lambda function, the startup time correspond to the time needed to start the function. For EMR, this correspond to the time of starting a cluster. The data files will be put in a S3 Bucket. The files will be streamed from the S3 bucket.

The Lambda function architecture is very simple. There is a Nodejs Server invoking each Lambda function asynchronously. Each Lambda function count the word of a chunk of a file. After counting the words of its chunk, the Lambda function write its result to a file. After invoking each functions, the server wait for them before aggregating all the results. When aggregating the results, the server read each file and merge it with the aggregator.

The implementation on EMR is fairly simple.  Apache Spark with the Scala driver is used to complete the benchmarked task. The following operations are executed:

\begin{enumerate}
\item Create a resilient distributed dataset (RDD) from s3 files
\item Read every lines to remove brackets with a regex
\item Split the lines in token using the space separator
\item Reduce to count the number of occurrence
\item Save the results back to S3
\end{enumerate}


%% file: 40.Results.tex
\section{\uppercase{Results}}
\label{sec:result}

\subsection{EMR results}

\noindent The costs were estimated based on the usage of each configuration and the execution time. For each configuration, there is one master node and n workers. Instances are of the type \textit{m4.large} and have an hourly rate of 13 cents. However, the usage is calculated based on the number of seconds. Table \ref{tab:emr_processing} presents the processing time results with different number of workers.

\begin{table}
\centering
\begin{tabular}{@{}cccc@{}}
\toprule
Number& Cluster& Processing& Cost\\
of workers&start-up time&time (minutes)&(dollars)\\ \midrule
2 & 3 & 75 & 0.507 \\
5 & 3 & 35 & 0.494 \\
10 & 3 & 20 & 0.548 \\ \bottomrule
\end{tabular}
\caption{Processing time and cost of different number of workers on EMR.}
\label{tab:emr_processing}
\end{table}

Unsurprisingly, increasing the number of workers reduces the processing time. Also, the processing time decrease linearly as more workers are added. This observation combined to the fact that the billing is based on the usage per seconds means that a cost effective and fast solution is to launch clusters with a high number of workers. This can be observed in results presented in  Table \ref{tab:emr_processing}. The cost of completing the task is smaller and the execution time faster when using 5 workers than when using 2. 

\begin{figure*}
    \centering
    \includegraphics[width=1\textwidth]{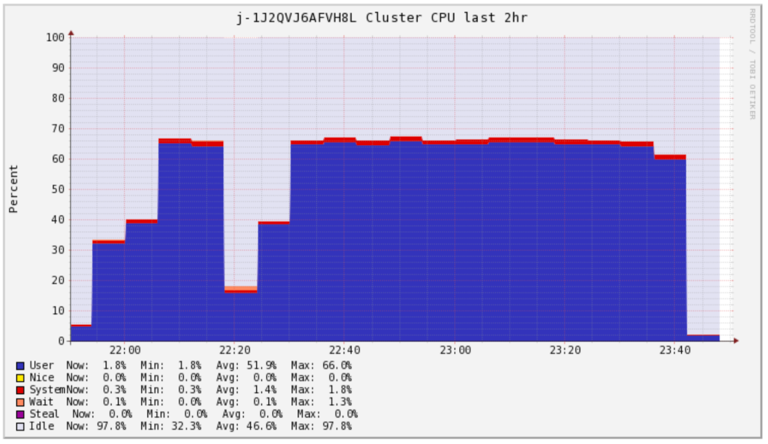}
    \caption{CPU load with 2 workers on Amazon EC2 cluster. This figure shows a constant cluster processing usage at 70\% starting at around 22:10. This first plateau correspond a preprocessing of the data done by Apache Spark. The mapReduce start at 22:30. The master node has almost no work which explains the ~70\% usage.}
    \label{fig:cpuload2workers}
\end{figure*}

\begin{figure*}
    \centering
    \includegraphics[width=1\textwidth]{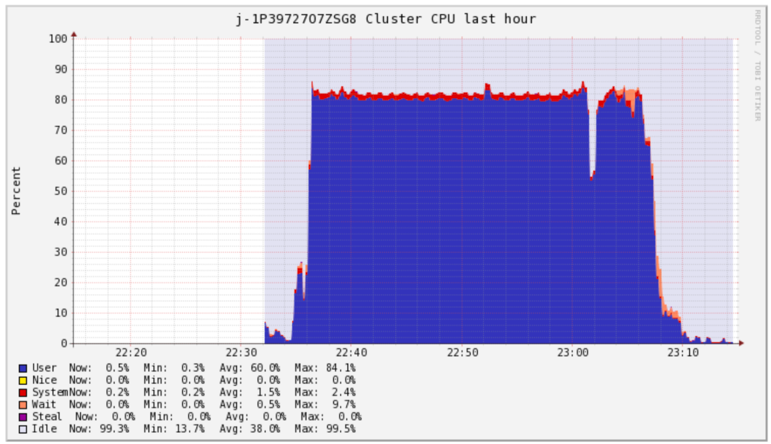}
    \caption{CPU load with 5 workers on Amazon EC2 cluster. This figure shows a constant cluster processing usage at 80\% starting at around 22:35. This first plateau correspond a preprocessing of the data done by Apache Spark. The mapReduce start at 23:05. Again, the master node has almost no work which explains the ~80\% usage.}
    \label{fig:cpuload5workers}
\end{figure*}

\begin{figure*}
    \centering
    \includegraphics[width=1\textwidth]{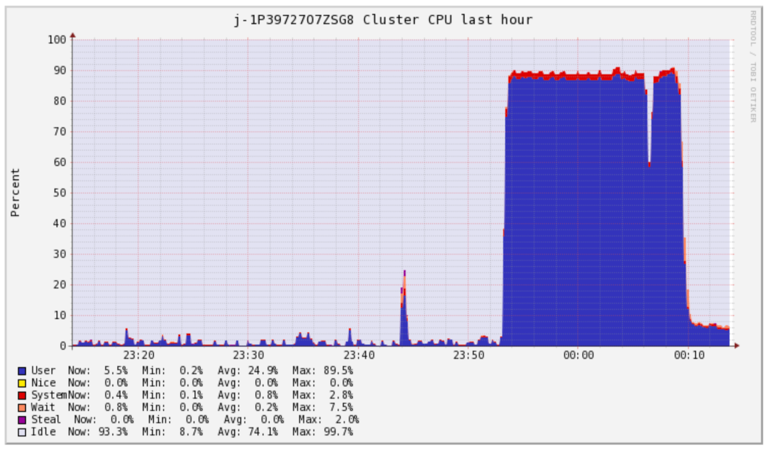}
    \caption{ CPU load with 10 workers on Amazon EC2 cluster. This figure shows a constant cluster processing usage at 90\% starting at around 23:52. This first plateau correspond a preprocessing of the data done by Apache Spark. The mapReduce start at 00:06. Again, the master node has almost no work which explains the ~90\% usage.}
    \label{fig:cpuload10workers}
\end{figure*}

A benefit of using AWS EMR is that just as by using Lambda, the data source is s3. As a result, the computation is separated from the data source (which is different from a classical Hadoop hdfs approach). The drawback is that the data has to be downloaded from s3. However, as we can observe in Figure \ref{fig:cpuload2workers}, Figure \ref{fig:cpuload5workers} and Figure \ref{fig:cpuload10workers}  this was not an issue. During the execution, all workers were working at maximum load which implies that network IO were not a bottleneck to the operations. 

\begin{figure*}
    \centering
    \includegraphics[width=1\textwidth]{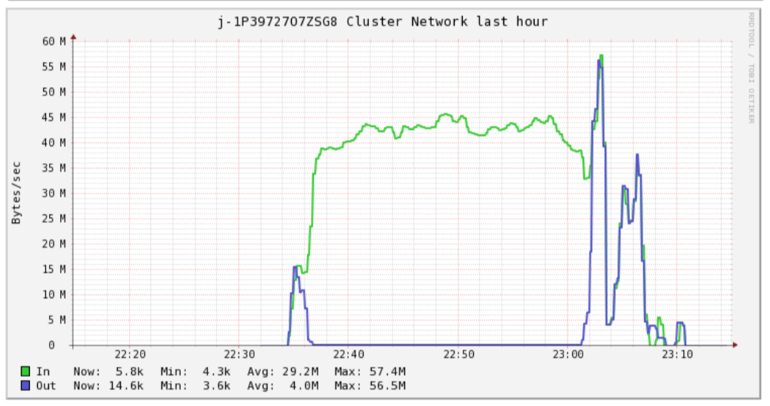}
    \caption{EMR Cluster Network with 5 workers. We can observe that the network throughput is initially going in the function. This is because EMR is requesting the data. After the processing, EMR write the results to S3 resulting to outbound throughput on the network.}
    \label{fig:Network5Workers}
\end{figure*}

\begin{figure*}
    \centering
    \includegraphics[width=1\textwidth]{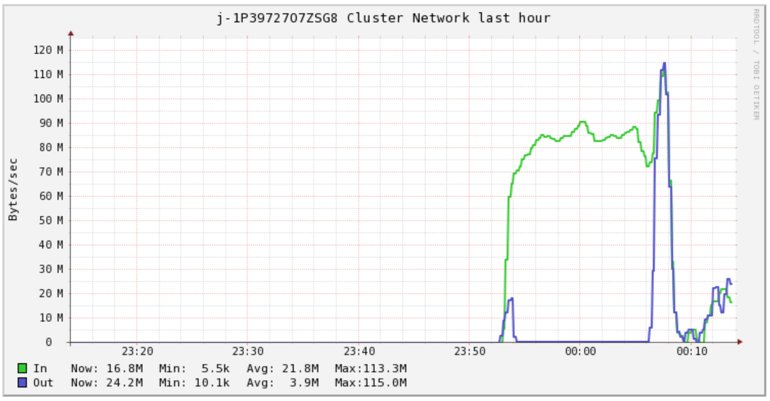}
    \caption{EMR Cluster Network with 10 workers. We can observe that the network throughput is initially going in the function. This is because EMR is requesting the data. After the processing, EMR write the results to S3 resulting to outbound throughput on the network.}
    \label{fig:Network10Workers}
\end{figure*}

The network throughput behaved as expected. During the execution, the data is being retrieved from s3 as it is processed which is observed from a nearly constant IN throughput. At the end of the execution, all the word counts are merged into a file that is sent to s3. This results in a burst in OUT throughput. The same behavior is observable in Figure \ref{fig:Network5Workers} and Figure \ref{fig:Network10Workers}.

\subsection{Lambda}

The costs were estimated based on the usage of each configuration and the execution time. The architecture required n workers that had n different execution times and one EC2 instance of type \textit{r4.xlarge} (hourly rate of 0.292 dollars). The average execution time of lambda is of 100 seconds. Instances were configured with 1.5gb of memory which implies that each execution (worker) cost 0.002501 dollars. presents the processing time results with different number of workers.

It is quite clear from the Table \ref{tab:awsTime.and.Cost} that the reduce section of the algorithm is the bottleneck of the process. However, the mapping part leveraging AWS lambda is fast. It should also be noted that the number of workers could be increased and just like EMR, the mapping section scales linearly.

\begin{table}
\centering
\begin{tabular}{@{}cccc@{}}
\toprule
Number&Map&Reduce&Cost \\
of workers&(minutes)&(minutes)&(dollars)  \\\midrule
89 & 3 & 100 & 0.202 + 0.486 \\ \bottomrule
\end{tabular}
\caption{Processing time and cost of the AWS Lambda pipeline.}
\label{tab:awsTime.and.Cost}
\end{table}

\subsection{Comparing Lambda function to Amazon Spark on EMR}

On table \ref{tab:comparingLambdaEMR}, we compare the results obtained using the AWS Lambda pipeline and the Amazon EMR pipeline. As we can see, the prices are in the same range. 

\begin{table}[h]
\centering
\begin{tabular}{@{}cccc@{}}
\toprule
Number&Cost EMR&Cost Lambda\\
of workers&(dollars)&(dollars)\\\midrule
89 & N/A & 0.686 \\ 
2&0.507&N/A\\
5&0.494&N/A\\
10&0.548&N/A\\\bottomrule
\end{tabular}
\caption{Cost comparison of the AWS Lambda pipeline and the Amazon Spark EMR.}
\label{tab:comparingLambdaEMR}
\end{table}

%% file: 50.Discussion.tex
\section{\uppercase{Discussion}}
\label{sec:discussion}

Both architectures will be compared based on three criteria: ease of use, cost and performance. Depending on the use case, some of these criteria would be more important than others. 

\paragraph{Ease of use} 
A considerable advantage of using EMR with Spark is that it is built to handle big data. As a result, programing the word count was trivial. Spark is doing most of the hard work of distributing data across different workers and optimize the calculation. The reduce section of the problem (merging the word counts) is not a problem for Spark because it uses a standard shuffling step to be able to perform the reduce step in a distributed way. In contrasts, the Node.js implementation requires that all the data be present in memory at the same time. This constraint was made possible by using an EC2 instance with more memory. However, with a bigger dataset or a different problem, this approach would not be suitable. Also, even if the mapping step was very efficient with lambda, the reduce step remains a bottleneck. 

The integration of Apache Zepellin also provided an ideal environment for experimentations.

\paragraph{Cost}
The cost per processing was lower for EMR. However, if the entire pipeline would be implemented in AWS Lambda, it might be different as only 0.22 dollars is billed for the current Lambda usage. 

\paragraph{Performance}
The processing time was significantly lower with the EMR pipeline. However, the mapping step of the Lambda pipeline was extremely fast. This suggests that other uses cases that requires a burst of parallel computing could be better suited for the Lambda architecture. By further increasing the number of workers, it could be used in real time application. In contrast, EMR has a too long start-up time to be launched on demand in a real-time scenario. It is therefore more adapted for long running offline tasks on large datasets.

%% file: 60.ThreatsOfValidity.tex
\section{\uppercase{Threats of Validity}}
\label{sec:threats}

\noindent The lambda implementation is suboptimal. In fact, it leverages distributed computation only for the map part of MapReduce. As a result, a significant bottleneck is present at the reduce step as it is performed on a single machine. Also, as a server is still necessary to perform the reduce step, the serverless advantages are less true.

The Apache Spark implementation could be further optimized by testing with different types of EC2 instances and configurations.

%% file: 80.Conclusion.tex
\section{\uppercase{Conclusion}}
\label{sec:conclusion}

\noindent In conclusion, we compared two services commonly available on most Cloud services platform. We looked at the classical Hadoop MapReduce framework with Apache Spark on Amazon EMR and at the newly available Serverless architecture with the Lambda Functions on AWS. 

We executed our benchmark on a typical MapReduce problem and showed that both techniques have similar performances. Thus, while doing our experiments, we learned that both tasks have really clear use case. We see that the EMR service should be used for very long compute task, whereas the Serverless compute is great for performing real time parallelized computing.